\documentclass[sigconf,screen,nonacm]{acmart}

\pdfoutput=1
\usepackage{todonotes}
\usepackage{esvect}
\usepackage{graphicx}
\usepackage{textcomp}
\usepackage{tabularx}
\usepackage{xcolor}
\usepackage{mwe}
\usepackage{subcaption}
\usepackage{caption}
\usepackage{xcolor}
\usepackage{algorithm}
\usepackage{algpseudocode}
\usepackage{xspace}
\usepackage{paralist}
\usepackage{booktabs}
\usepackage{multirow}

\newcolumntype{P}[1]{>{\RaggedRight\hspace{0pt}}p{#1}}
\def\BibTeX{{\rm B\kern-.05em{\sc i\kern-.025em b}\kern-.08em
    T\kern-.1667em\lower.7ex\hbox{E}\kern-.125emX}}

\definecolor{belizehole}{HTML}{2980b9}
\definecolor{clouds}{HTML}{f4f8f9}
\definecolor{midnightblue}{HTML}{2c3e50}
\definecolor{pomegranate}{HTML}{c0392b}
\definecolor{royalblue}{HTML}{3867d6}
\definecolor{greensea}{HTML}{16a085}
\definecolor{nephritis}{HTML}{27ae60}
\definecolor{amethyst}{HTML}{9b59b6}

\usepackage{listings}
\AtBeginDocument{\DeclareCaptionSubType{lstlisting}}
\usepackage{subcaption}

\usepackage[frozencache,cachedir=.]{minted}

\usepackage{enumitem}
\AtBeginDocument{%
  \providecommand\BibTeX{{%
    Bib\TeX}}}

\def\hpacgpu{\textsc{hpac-offload}\xspace}
\begin{document}

\title{HPAC-Offload: Accelerating HPC Applications with Portable Approximate Computing on the GPU}

\author{Zane Fink$^1$, Konstantinos Parasyris$^2$, Giorgis Georgakoudis$^2$, Harshitha Menon$^2$ \\
$^1$University of Illinois at Urbana-Champaign, Champaign, Illinois, USA \\
zanef2@illinois.edu\\
$^2$ Lawrence Livermore National Laboratory, Livermore, CA, USA \\
\{parasyris1, georgakoudis1, harshitha\}@llnl.gov
}

\renewcommand{\shortauthors}{Fink et al.}

\begin{abstract}

The end of Dennard scaling and the slowdown of Moore's law led to a shift in technology trends towards parallel architectures, particularly in HPC systems. To continue providing performance benefits, HPC should embrace Approximate Computing (AC), which trades application quality loss for improved performance. However, existing AC techniques have not been extensively applied and evaluated in state-of-the-art hardware architectures such as GPUs, the primary execution vehicle for HPC applications today.

This paper presents HPAC-Offload, a pragma-based programming model that extends OpenMP offload applications to support AC techniques, allowing portable approximations across different GPU architectures. We conduct a comprehensive performance analysis of HPAC-Offload across GPU-accelerated HPC applications, revealing that AC techniques can significantly accelerate HPC applications (1.64x LULESH on AMD, 1.57x NVIDIA) with minimal quality loss (0.1\%). Our analysis offers deep insights into the performance of GPU-based AC that guide the future development of AC algorithms and systems for these architectures.

\end{abstract}

\maketitle

\section{Introduction}
As Dennard scaling --- which stipulated a steady rise in processor clock speed through transistor shrinkage --- came to an end, and Moore's law --- predicting a doubling of CMOS transistors on a microchip every two years --- slowed down, technology trends shifted toward parallel architectures. Parallel architectures focused on multi-core CPUs in the early 2000s, while the emergence of GPGPU paradigms pivoted technology trends to many-core accelerator systems.
This trend is evident in the Top500 list~\cite{top500list}: as of November 2022, $7$ of the $10$ fastest supercomputers use GPUs. Despite the success of many-core architectures overcoming the slowdown of Moore's law ~\cite{hennessy2019new}, HPC requires another paradigm shift to continue delivering performance improvements.

Approximate Computing (AC) has emerged as an attractive new paradigm that increases performance by introducing novel approximations within applications, controllably reducing the application's accuracy. Both hardware and software AC techniques have been proposed. Specifically,~\cite{froehlich2018towards, rehman2016architectural} introduce approximate CPUs, ~\cite{esmaeilzadeh2012architecture} proposes approximate memories while ~\cite{esmaeilzadeh2012neural, grigorian2015brainiac} discuss approximate accelerators.
Software techniques include loop perforation~\cite{Hoffmann2009, sidiroglou2011managing}, which accelerates image processing workloads by up to $3\times$ with less than $10\%$ accuracy loss. Input~\cite{mishra2014iact}  and output~\cite{tziantzioulis2018temporal} approximate memoization have been used in various domains, such as stencil computations, finance, and image processing, doubling application performance with small error. Other techniques, such as variable precision, can increase performance by $45\%$. HPAC~\cite{parasyris2021hpac} provides a state-of-the-art compiler and runtime implementation to apply software AC techniques on multi-core CPUs using OpenMP.

These works extensively showcase the potential of AC in various CPU applications. However, little research assesses approximate computing on GPUs, which currently dominate HPC supercomputers. It is imperative to assess whether AC is a viable execution paradigm for next-generation software: any paradigm that cannot apply to many-core architectures will likely have limited impact. Consequently, a comprehensive study applying AC to GPU-enabled applications is essential to fully gauge the potential and challenges imposed by approximations in modern GPUs.

To address this problem, this work studies state-of-the-art software approximate computing techniques applied to HPC GPU-enabled scientific applications. We present \hpacgpu{}, an extension of HPAC~\cite{parasyris2021hpac} that supports approximations in GPU applications. The proposed extensions seamlessly compose with the portable OpenMP offload programming model and consist of easy-to-use annotations on OpenMP offload applications. The result is an approximate computing framework that enables portable approximations across different GPU architectures, such as NVIDIA and AMD. The composition of approximation and GPU parallel execution results in several challenges due to the execution model of GPU devices. \emph{Porting AC techniques to GPUs without considering their unique architectural characteristics results in significant slowdowns.}

For example, approximate computing techniques for CPU parallelism typically duplicate the AC state on each CPU thread; however, the massive parallelism of GPUs that use millions of software threads makes this approach impractical by depleting the device's memory. Additionally, CPU-AC allows each parallel thread to independently decide whether to approximate without observable overhead. In contrast, independent thread decision-making in GPUs can introduce thread divergence and reduce performance, limiting the expected performance boosts of approximation.

As such, programming models for GPU-AC must match the hierarchical nature of the underlying execution model. \hpacgpu{} identifies such challenges and proposes programming model extensions to support high-performance implementations of several state-of-the-art approximation techniques: input/output memoization and loop perforation.

This paper makes the following contributions:
\begin{itemize}[leftmargin=*]
    \item \hpacgpu{}, a programming model for composing state-of-the-art AC techniques (input/output memoization, loop perforation) with OpenMP-offload. Our pragma-based programming model equips developers with AC techniques using familiar idioms. \hpacgpu{} reflects the hierarchical programming model of modern GPUs, enabling users to apply AC with little effort.%
    \item New GPU-centric algorithms for approximate computing techniques that leverage the architectural features of modern GPUs. Our algorithms offer hierarchical control of approximation, coupling flexibility with high performance.
    \item An implementation of \hpacgpu{} in Clang/LLVM. \hpacgpu{} leverages the portable OpenMP offload specification and extends it with a portable layer to apply approximations to GPU-enabled applications. %
    \item Evaluation on a comprehensive suite of proxy apps, mini-apps, and benchmarks from HPC and machine learning on AMD and NVIDIA GPUs. We demonstrate that \hpacgpu{} achieves up to $6.9\times$ speedup across all applications (geomean speedup $1.42\times$) while introducing less than $10\%$ error.
\end{itemize}

\section{Background}
\label{sec:overview}
This section provides background information on GPU architectures and their execution model, the OpenMP offload programming model, and the HPAC programming model.
We use CUDA~\cite{cuda2021guide} terminology; other GPU vendors have similar features.

\subsection{GPU Architecture and Execution Model}
A GPU contains processing units called streaming multiprocessors (SM), each with multiple cores, a register file, an L1 cache, shared memory, and a read-only cache. GPUs follow a Single Program, Multiple Data (SPMD) interface, where GPU threads run the same kernel. All threads in the kernel are organized into a grid. This grid is divided into parallel threads grouped into thread blocks, which are further divided into warps. Within a warp, threads use a Single Instruction Multiple Data (SIMD) execution paradigm. Register files are thread-private memory, while other memory types are shared. SMs access global memory through high-bandwidth, high-latency memory transactions. Larger transactions reduce DRAM operations, and memory coalescing combines thread requests when the requested memory addresses fit within a transaction.

In the GPU-accelerated execution model, time-consuming tasks are offloaded as kernels to the GPU through kernel offloading. The host transfers data between Host and Device memory (HtoD and DtoH) and invokes the kernel.

\subsection{OpenMP Offload}
OpenMP offload is a set of pragma directives first introduced in OpenMP 4.0 
that allows offloading execution to heterogeneous devices such as GPUs. 
The OpenMP offload model uses the \verb|target| directive to define offloading for a code region (target region) to a device, along with a data mapping. The mapping uses modifiers like \verb|to|, \verb|from|, and \verb|tofrom| to specify data directionality. The target directive executes the region on a single device thread.

The \verb|teams| directive forms a league of teams, with each team's main thread executing the region. The \verb|parallel| directive allows all threads in a team to execute the region. In CUDA, a team represents a thread block, and the \verb|parallel| directive assigns a kernel to all threads in the block. The \verb|#pragma omp target teams parallel| construct, combined with work-sharing directives (\verb|distribute| and \verb|for|), decomposes parallel loops across teams and threads.

Figure \ref{fig:ompExample} shows an OpenMP offload example. The algorithm applies the \verb|foo| function to all elements of an input vector, storing the result in an output vector. The developer declares the \verb|foo| function as a device function using the \verb|omp declare target| directive (lines 1, 3), directing the compiler to include it in the device binary. The developer also implements a parallel loop for device execution using the \verb|pragma omp target teams distribute| \verb|parallel for| (line 7). The \verb|teams distribute| directive shares iterations across thread blocks, while the \verb|parallel| directive divides iterations among the threads of each block.  The \verb|map| directive (line 8) copies the \verb|input| vector to the device and the \verb|output| vector from the device.

\begin{figure}
\centering
\begin{minted}[numbersep=1pt,fontsize=\footnotesize,linenos,frame=lines,escapeinside=!!]{c++}
#pragma omp declare target 
double foo(double input) { ... }; // An expensive function
#pragma omp end declare target 

void Hfoo(double *input, double *output, size_t N) 
{
  #pragma omp target teams distribute parallel for \
   map(to: input[0:N]) map(from: output[0:N])
  for(size_t i = 0; i < N; ++i)
    output[i] = foo(input[i]);
}
\end{minted}
\caption{An OpenMP-offload example.}
\label{fig:ompExample}
\end{figure}

\subsection{State-of-the-art of AC}\label{sub:sota}

Many approximate computing techniques are specific to one domain, including mixed-precision~\cite{chiang2017rigorous}, Newton's method~\cite{Galantai2000:Theory}, and the Barnes-Hut algorithm~\cite{Barnes1986:Hierarchical}. These techniques enhance performance but require domain expertise and familiarity with the application code.
This work, however, concentrates on general-purpose AC techniques applicable to multiple domains. We examine three state-of-the-art software-based methods: \emph{loop perforation}~\cite{misailovic2011probabilistically,misailovic2010quality}, \emph{input memoization} (iACT~\cite{mishra2014iact}), 
and \emph{output memoization} (TAF~\cite{tziantzioulis2018temporal}).
After briefly presenting these techniques, we describe their adaptation to GPU architectures and evaluate their impact on accuracy and performance in scientific applications.

\textbf{Output Memoization (TAF)} leverages temporal function locality: recent past outputs resemble upcoming ones. Instead of computing upcoming evaluations, TAF returns the most recent evaluation's output.
TAF caches code region outputs in a sliding window of history size (\emph{hSize}) output values, calculating the relative standard deviation  ($RSD$)\footnote{Relative standard deviation is given by $RSD = \sigma / \mu$ for population standard deviation $\sigma$ and population mean $\mu$.}. When the $RSD$ value is below a user-defined threshold, TAF enters a stable
regime for the next  prediction size (\emph{pSize}) invocations (\emph{hSize} and \emph{pSize} are user-specified), returning the last accurately-computed output. TAF employs a state machine at runtime to track the current sliding window's outputs and determine if it is in a stable regime and should approximate.

\textbf{Approximate input memoization (iACT)} extends traditional memoization (caching). iACT caches the inputs and outputs from code region evaluations. For every new evaluation, the technique computes
the distance of the inputs in the cache with the inputs of the current evaluation. When the distance is below
some user-defined threshold, input memoization returns the closest previously computed value, skipping the computation of the code region.

\textbf{Loop perforation} is an approximation technique that 
drops the computations of user-specified iterations. Typically, the user describes
which iterations to drop through some pattern. For example, the user can 
skip one of every $M$ iterations (\emph{small} perforation) or execute one of every $M$ iterations (\emph{large} perforation). Other perforation types  
include \emph{ini} and \emph{fini}, which drop a user-defined fraction of the first or last loop iterations.

\begin{figure}
\centering
\begin{minted}[numbersep=1pt,fontsize=\footnotesize,linenos,frame=lines,escapeinside=!!]{c++}
double foo(double input) { ... }; // An expensive function

void Hfoo(double *input, double *output, size_t N) {
#pragma approx perfo(small:4)
#pragma omp parallel for
  for(size_t i = 0; i < N; ++i)
#pragma approx memo(in: 10 : 0.5f) in(input[i]) \
                out(output[i])
    output[i] = foo(input[i]);
}
\end{minted}
\caption{An example of the HPAC programming model.}
\label{fig:HPACExample}
\end{figure}

The \textbf{HPAC programming model}~\cite{parasyris2021hpac} enables pragma-based approximate computing, integrating OpenMP parallelism with approximate computing techniques such as input/output memoization and loop perforation. Developers can use HPAC's toolchain to identify code regions amenable to approximate computing and evaluate accuracy/performance trade-offs.

\textbf{HPAC example.} Figure~\ref{fig:HPACExample} shows an example composing the HPAC programming model with
CPU OpenMP. The developer applies
\emph{small} perforation to the \verb|parallel for| to skip the computation of
every fourth iteration in every parallel thread (line 4). The developer applies input memoization for the remaining executed iterations using a cache table of size $10$ (line 7). Memoization is activated when the euclidean distance between the current input \verb|input| and a cache entry is less than \verb|0.5f|. If such an entry exists, the memoization cache returns the output value associated with this entry instead of calling the \verb|foo| function. The cache size and the distance threshold are parameters to the \verb|memo| clause.

\textbf{Design of HPAC.}
The HPAC \verb|pragma|s support approximate computing by automatically creating a second \emph{approximate execution path} that co-exists with the original \emph{accurate execution path} during compilation and execution. Developers specify the desired approximation technique through HPAC clauses, and HPAC creates the approximate execution path with the corresponding technique's implementation.
At execution time, the HPAC runtime library decides whether to execute the \emph{accurate} or \emph{approximate} execution path. For each supported approximation technique, HPAC provides a parametric \emph{activation function} that dynamically determines which execution path to follow. If the accurate path is chosen, the code region is executed, and inputs/outputs are captured if necessary. When the approximate path is selected, the AC technique approximates the output.

The HPAC execution harness exhaustively explores the space of user-provided approximation techniques (e.g., iACT and TAF) and parameters (such as error threshold). The technique and parameters are first applied to the program by the harness, which builds and executes the program. After executing the approximated program, the harness calculates and saves runtime information and error to a database. Using these data, the user can decide how approximation fits her application needs.

\section{Design and Implementation}

This section details the  design and implementation of \hpacgpu{}. While we focus on the AC techniques introduced by the original HPAC work, our implementation is flexible and extensible to accommodate future AC techniques. 

GPUs combine large numbers of threads with rapid context switching between them to hide execution and memory latencies. The deep memory hierarchy, memory coalescing, and SIMD execution model reduce memory latency and enable massive parallelism. An AC framework oblivious to these characteristics will not improve performance. %

\subsection{GPU Aware AC Design}\label{sec:challenges}

\subsubsection{AC GPU memory design}\label{subsub:memory}
Memory-bound GPU applications present a significant challenge for GPU AC. %
In memory-bound applications, 
memory bandwidth determines performance. Performance will be harmed by AC techniques that cache results in memory to avoid computation but increase memory traffic. %

\begin{figure}
     \centering
      \includegraphics[width=0.35\textwidth]{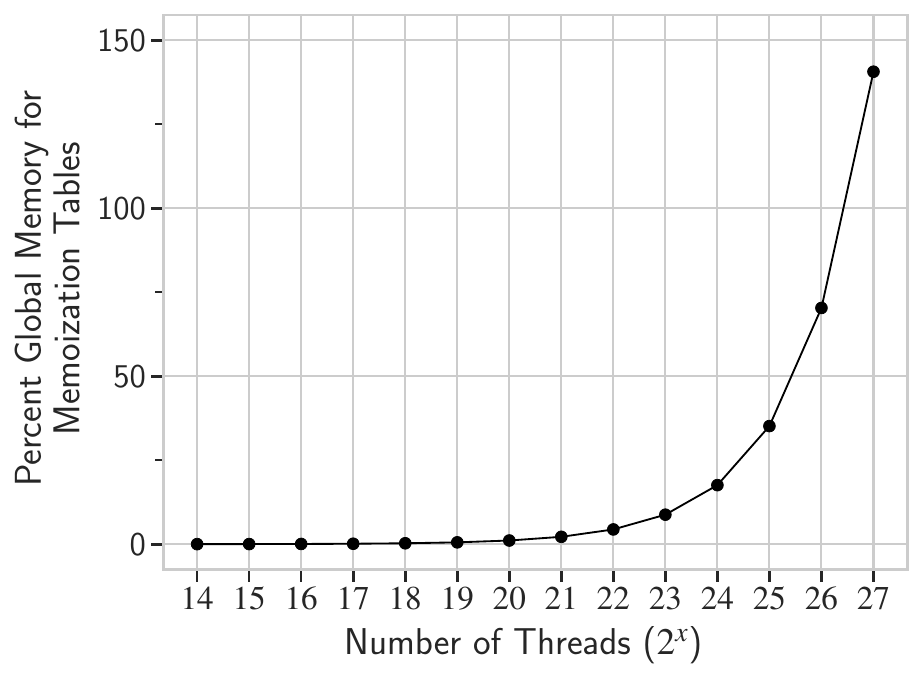}
        \caption[]{Number of threads (x-axis) vs. the percent of an NVIDIA V100 GPU's 16GB of global memory needed to store per-thread memoization tables (y-axis). We assume 
        a memoization cache table of 5 entries; each entry is of size 36 bytes.}
        \label{fig:global_mem_impact}
\end{figure}

The resources dedicated to each parallel thread in GPU execution are far more constrained than the respective CPU resources. AC techniques unaware of such limitations will starve the system of resources. For example, in approximate memoization, each thread replicates internal AC data structures (e.g., the memoization table). Adopting this on the GPU is unviable, as depicted by Figure~\ref{fig:global_mem_impact}. The AC data structures fill the device's global memory when the application uses $2^{27}$ threads, far below the limit of $\approx 2^{72}$~\cite{cuda2021guide}.

AC techniques that are agnostic to these hardware characteristics will not improve performance.
\emph{\hpacgpu{} employs a GPU-aware memory scheme to facilitate efficient AC for GPUs.}

To reduce GPU AC's excessive resource consumption, we rely on the following key observation: \emph{while a kernel may have millions of threads, 
a small subset of those threads are actively scheduled on SMs at a given time.}
Threads scheduled on the same SM over non-overlapping time periods can 
use the same memory to store internal AC state and thus 
substantially reduce storage requirements. We move all internal AC state to a block's shared memory to avoid device-wide locking. Since the number of
active threads is bounded by the hardware specification (number of SMs, number of active threads per
SM) and orders of magnitude smaller than the total number of supported threads, \hpacgpu{} substantially reduces the resources used to store the AC internal state.

Storing the AC state in shared memory facilitates fast, local access to 
internal AC data structures. However, shared memory data only persist during the lifetime
of the currently executing kernel, and \hpacgpu{} approximations are thus scoped within the
kernel lifetime. Once the kernel completes, the internal data are destroyed.
\emph{In summary, \hpacgpu{} limits resource consumption by dedicating part of the shared 
SM memory to the internal AC state to leverage the GPU memory hierarchy for performance.}

\subsubsection{Prescriptive and Hierarchical GPU AC}\label{subsub:hierarchical}

GPU applications decompose their computation into blocks of threads, which are further decomposed into warps that consist of individual threads. These applications frequently use fine-grain synchronization and warp level 
collective primitives (shuffle intrinsics) to provide high-performance implementations of complex algorithms.

The data-dependent nature of activation functions can introduce deadlocks within the application.
Approximation skips the accurate execution path, which can lead to a deadlock when the accurate path contains synchronization primitives.
Because of the SIMD execution model, activation functions can also introduce divergence in GPU applications. 
In a worst-case scenario, one thread in the warp takes the accurate execution path while all others approximate. Thus, the entire warp will wait for this single thread to finish executing, reducing the performance gains of approximation.

Such scenarios introduce deadlock and accuracy loss while limiting the broad adoption and performance benefits of AC.
\emph{\hpacgpu{} provides programming model extensions to developers to define the hierarchical level
of approximation.}

\hpacgpu{} considers three hierarchy groups: ($i$) \emph{thread-level}: where each thread has individual approximation criteria; ($ii$) \emph{warp-level}: all threads in a warp share collective criteria; and ($iii$) \emph{block-level}: all threads in a block share collective criteria. 
When the collective criteria are satisfied, the entire hierarchy group approximates; otherwise, all threads follow the accurate execution path.

\subsubsection{Temporal Approximate Function Memoization (TAF)}
\begin{figure*}
    \centering
    \includegraphics[width=0.8\textwidth]{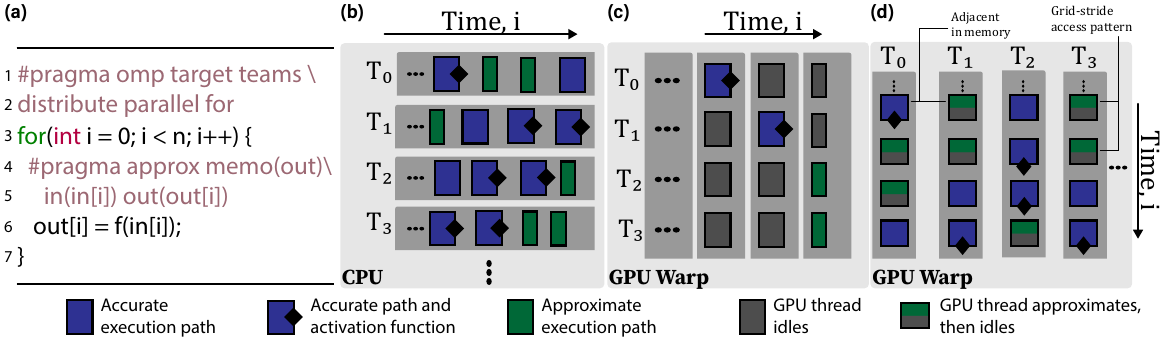}
    \caption{Different algorithmic TAF implementations using $2$ for the prediction and history sizes for the parallel loop shown in Panel (a) on CPU and GPU architectures. CPU threads execute contiguous, non-overlapping loop iterations; adjacent GPU threads execute adjacent iterations. In (b), each CPU thread independently manages TAF state and controls approximation using the activation function. (c) The semantically-equivalent GPU version fulfills TAF's spatial locality assumption, but GPU threads remain idle awaiting activation criteria fulfillment. (d) shows the \hpacgpu{} TAF algorithm, where GPU threads independently manage TAF state, but spatial locality is violated.}
    \label{fig:taf_gpu}
\end{figure*}

TAF assumes that a thread runs the same function in sequence with outputs that exhibit spatial locality over time.
The history size (\emph{hSize}) previous outputs determine whether the accurate or approximate execution paths are taken.
The structure adds control dependencies between previous loop iterations and the current iteration.
HPAC, as illustrated in Figure~\ref{fig:taf_gpu}(b), adopts the same design. In CPU parallel for loops, a thread 
executes adjacent \verb|for| iterations to preserve cache locality, and therefore parallel CPU-based
TAF is simple and maintains the semantics of temporal output memoization.

On the other hand, \verb|target parallel for| loops are distributed among adjacent threads executing adjacent iterations.
In a semantically-equivalent TAF algorithm for the GPU, threads must wait for the previous thread to terminate
before deciding whether to approximate, serializing execution (Figure~\ref{fig:taf_gpu}(c)).
To eliminate this serialization, we relax TAF's spatial locality assumption in \hpacgpu{}'s TAF algorithm. %

The \hpacgpu{} TAF algorithm is depicted in Figure~\ref{fig:taf_gpu}(d), where
threads exhibit temporal output locality without requiring spatial locality.
In \verb|target parallel for| loops, locality manifests across the iterations 
of a grid-stride step. With this assumption, no inter-thread dependencies are introduced, but utilization is limited by divergence-induced idle time. Nevertheless, parallelism is increased and coalesced memory accesses are preserved, yielding a high-performance TAF algorithm for the GPU.

\subsubsection{Input Approximate Function Memoization (iACT)}\label{subsub:iact_impl}

iACT presents interesting trade-offs. 
Larger memoization tables increase approximation likelihood~\cite{parasyris2021hpac}, search costs, and memory use.
CPU-HPAC allocates unique memoization tables per thread, minimizing synchronization but increasing memory overhead. To decrease memory overhead on the GPU, this scheme is adjusted to allow \emph{iACT memoization tables to be shared within a warp}.

 Developers control whether threads share memoization tables and, if so, the number of threads that share each table. Table sharing allows us to mediate memory overhead, serialization, and synchronization with the benefits of a larger aggregate table size.
 This extension has a few advantages: 
 \begin{inparaenum} 
 \item warp-level sharing reduces synchronization, requiring coherence only among threads sharing a table;
 \item warp-level sharing allows threads to access computed values from adjacent threads, boosting approximation rates;
 \item balancing memory use and synchronization/serialization overhead is possible through sub-warp table sharing. Private per-thread tables increase parallelism \emph{and} memory use. Conversely, sharing one table per warp reduces memory use but requires synchronized access to prevent data races. Our design allows us to explore these trade-offs.
 \end{inparaenum}

\subsubsection{GPU-Aware Loop Perforation} Loop perforation skips a subset of a loop's iterations according to the method used. On the GPU, small and large perforation suffer performance degradation because neighboring threads in a warp take separate execution paths, introducing thread divergence. To eliminate thread divergence, we introduce \emph{herded perforation}. In herded perforation, the same iterations are dropped by every thread in the grid. For instance, if we skip every third iteration, all threads in the grid will skip the respective iterations. This synchronized skipping ensures that the control flow remains uniform across all threads in a warp, thereby eliminating thread divergence. Maintaining the same perforation pattern throughout the warp reduces global memory traffic, as the memory transactions are aligned and less fragmented. Furthermore, deadlock is avoided because threads in the block take the same execution path.

\begin{figure}
\centering
\begin{minted}[numbersep=1pt,fontsize=\footnotesize,linenos,frame=lines,escapeinside=!!]{c++}
#pragma omp declare target // Expensive device functions
double foo(double *input, int nDims, int rowsize) { ... };
double bar(double *input) { ... };
#pragma omp end declare target 
void Hfoo(double *input, double *output, size_t N) {
  #pragma omp target teams distribute parallel for \
   map(to:input[0:5*N]) map(from:output1[0:N], output2[0:N])
  for(size_t i = 0; i < N; ++i){
    #pragma approx memo(in:2:0.5f:4) level(warp) \
    in(input[i*5:5:N]) out(output1[i])
    output[i] = foo(&input[5*i], 5, N);

    #pragma approx memo(out:3:5:1.5f) level(thread) \
    out(output2[i])
    output2[i] = bar(&input[i]);
  }
}
\end{minted}
\caption{\hpacgpu Example.}
\label{fig:hpacoffload_example}
\end{figure}

\subsection{Programming Model}\label{sub:programming_model}
\hpacgpu uses non-intrusive pragma-based annotations for approximation, allowing AC with minimal modifications to existing code. 
To support the discussed extensions, we introduce a new clause \verb|level(hierarchy)| that can be applied to the original HPAC \verb|approx| directive.
This new optional clause determines the hierarchy level in which approximation will be applied. Allowed values are \verb|thread|, \verb|warp|, \verb|team|.
The default value is \verb|thread|, closely matching the original HPAC programming model. When \verb|warp| or \verb|team| is used,
threads collectively decide which execution path to follow.

Further, we extend the input memoization clause to accept an additional optional parameter \verb|tperwarp| defining the number
of tables per warp (\verb|memo(in:tsize:threshold:tperwarp)|). The warp size is the default value, yielding one independent table for each thread.
Lower values increase sharing among threads, increasing synchronization overhead but also reducing shared memory use.

A program using \hpacgpu is shown in Figure~\ref{fig:hpacoffload_example}. Using the pragma-based programming model, users choose and parameterize an approximation technique for a code region. To approximate a code region, the developer first chooses which category of AC technique to use (e.g., perforation or memoization), then specifies the particular type and supplies parameters to the technique.

Memoization is specified with the \texttt{approx memo} clause, followed by a keyword designating the memoization algorithm: either \texttt{in} for iACT or \texttt{out} for TAF. The keyword is followed by a colon-separated argument list. In line $9$ of Figure~\ref{fig:hpacoffload_example}, the developer specifies iACT with a table size of 2, a distance threshold of \verb|0.5f|, and 4 tables per warp. Warp-level decision-making is specified via \verb|level(warp)|. In line $10$, she declares the approximated region's inputs and outputs with the array section \verb|input[i*5:5:N]|. The array is a $5$-dimensional vector stored in column-major format to maximize memory coalescing, so strided memory access is used with stride $N$.

A second function is approximated using TAF in lines 13-15 of Figure~\ref{fig:hpacoffload_example}. In line 13, TAF is selected with a history size of 3, prediction size 5, and RSD threshold $1.5f$. Thread-level TAF is selected via \verb|level(thread)|. TAF only uses a code region's outputs, so no input is declared in line 14.

\subsection{Implementation}\label{sec:taf_impl}
We initially rebased HPAC to Clang/LLVM~\cite{lattner2008llvm} version 16 and modified it with the extensions necessary for \hpacgpu.
We extend the parser to recognize the new clause (\verb|level|) and the new identifiers defining the tables per warp. This information is lowered to the semantic analysis and embedded in an AST node for the approximation directive. The annotated code region is captured as a closure in Clang, making the accurate, non-approximate version callable as a function. We extend the device code generation for the AST node to allocate and initialize several data structures with the information needed to control and perform the approximation technique. The compiler generates a call to the runtime function whose arguments have the information needed to perform the approximation. 
The \hpacgpu{} runtime system is implemented as OpenMP offload device functions, making it portable to all architectures supported by the OpenMP offload implementation. 

To reduce the storage requirements of GPU AC, the internal AC state is stored in shared memory (Section~\ref{subsub:memory}). The amount of needed shared memory is given by the user when building the \hpacgpu runtime library.\footnote{The Clang/LLVM OpenMP implementation lacks full support for dynamic shared memory. When full support is added, we will update \hpacgpu accordingly.} At runtime, TAF and iACT use this shared memory.

In iACT, we enable table sharing by dividing table access into reading and writing phases. Threads search for input matches during the reading phase, while a single writer is chosen for each table during the writing phase. The writer is the thread with the largest euclidean distance from any table value. A warp barrier separates phases, and we use a round-robin replacement policy.\footnote{We also implemented CLOCK~\cite{Corbato1968:Paging} and found no effect.}

TAF is implemented in the \hpacgpu runtime system as shown in Figure~\ref{fig:taf_gpu}(d). Each thread manages its private shared-memory TAF state machine and output memoization table. RSD is calculated over the outputs of successive runs of the accurate execution path over a grid-stride loop. When in the approximate state, a thread writes its newest table value to the region's output. When in the accurate state, a thread executes the accurate execution path and copies the output to the thread's output table.

To support \emph{small} and \emph{large} perforation at runtime, \hpacgpu counts the number of times a thread has encountered the perforated code region. A thread skips the code region according to the perforation technique used. For \emph{ini} and \emph{fini} perforation, the compiler generates code to change the lower or upper bounds of the loop, respectively. Perforation can be applied to both \verb|omp| \verb|target teams distribute| loops and loops within offload kernels.

To support hierarchical decision-making, threads in a hierarchy group tally the threads whose activation criteria have been met. Using a ``majority-rules'' system, the entire group approximates if most of its threads meet the activation criteria. For warp-level decision-making, the \verb|ballot| intrinsic identifies threads that will approximate; \verb|popcount| counts these threads. For block-level decision-making, threads are counted separately in each warp using \verb|ballot| and \verb|popcount|. The first thread in each warp atomically adds its count to the block total in shared memory, and the total is read after all warps contribute. If the majority of threads can approximate, the entire block follows suit. Although we focus on the ``majority-rules'' scheme, \hpacgpu{} can easily be extended to support others.

\section{Evaluation}
\label{sec:evaluation}

We evaluate the approximation techniques implemented in \hpacgpu using the benchmarks listed in Table~\ref{tab:benchmarks}. We first profile each benchmark to find the longest-running offload kernel and decorate code regions within each kernel with approximation pragmas. If the two longest-running offload kernels have similar runtime, we approximate them both. We try to maximize the approximated portion of the kernel to increase potential performance benefits. 

To explore accuracy, performance trade-offs and GPU parallelism, we perform a design space exploration of \hpacgpu{} parameters
and the \verb|num_teams| OpenMP clause parameter. By adjusting the value passed to \verb|num_teams|\footnote{While we vary num\_teams, we use the one value of num\_threads that yields the best performance in the non-approximated benchmark.}, we can assign
more items to be computed by the same GPU thread and thus explore the interaction between parallelism and approximation. %
Table~\ref{tab:params} 
lists all parameter values, and we explore the Cartesian product of these parameters.

Except for Blackscholes\footnote{We perform 8 trials for each Blackscholes configuration to get significant results.}, we run each configuration 3 times and report the mean runtime. These parameters yield statistically significant results. Unless otherwise stated, we measure the end-to-end application runtime, including time transferring data between the CPU and GPU, and report speedup relative to this end-to-end runtime. To reduce overplotting, we divide the error range for each benchmark into ten equally-sized intervals. For each interval, we show the fastest and slowest $10\%$ of configurations.

We showcase \hpacgpu{} portability by evaluating it on two state-of-the-art GPU platforms. 
The first is equipped with $2\times$ IBM Power9 CPUs, each with $22$ cores, and $4\times$ NVIDIA Tesla V100 GPUs, each with 80 SMs; the other 
has $1\times$ AMD Epyc 7A53 CPU with $64$ cores and $4\times$ AMD Instinct MI250x GPUs, each with 220 SMs.

We evaluate \hpacgpu{} on two metrics: end-to-end speedup over the original GPU-accelerated application, and quality loss. 
To quantify quality loss, we choose a Quantity of Interest (QoI) from each application and report approximation-induced error 
in comparison to the output of the original application. 
As an error metric we use  mean absolute percent error (MAPE) (\ref{eq:mape}) for all applications besides K-Means for which we use the misclassification rate (\ref{eq:mcr}) (MCR). 
\begin{equation}\label{eq:mape}
        MAPE(\vv{O_{ac}}, \vv{O_{ap}}) = \frac{1}{N} \sum_{i=1}^N \frac{\lvert \vv{O_{ac}}(i) - \vv{O_{ap}}(i) \rvert}{\vv{O_{ac}}(i)}
\end{equation}

\begin{equation}\label{eq:mcr}
        MCR(\vv{O_{ac}}, \vv{O_{ap}}) = \frac{1}{N}\sum_{i=1}^N{\mathcal{I}[\vv{O_{ac}}(i) \ne \vv{O_{ap}}(i)}]
\end{equation}

$\vv{O_{ac}}$ and $\vv{O_{ap}}$ represent the outputs of the accurate and approximate executions, respectively. $\mathcal{I}[x]$ is the indicator function that returns $1$ if and only if $x$ evaluates to true.

\begin{table}
\caption{The benchmarks used to evaluate \hpacgpu{}.}
\scriptsize
\begin{tabular}{p{0.15\columnwidth} p{0.81\columnwidth}}
\toprule
\textbf{Benchmark} & \textbf{Description}   \\
\midrule
LULESH~\cite{Karlin2013:LULESH}             & Hydrodynamics proxy application that models a Sedov blast problem with volumetric elements discretized onto a mesh. {\bfseries QOI:} The final origin energy. \\ \midrule
Leukocyte~\cite{Rodinia}          & Detects and tracks rolling white blood cells (leukocytes) in video microscopy of blood cells.  {\bfseries QOI:} The final location of each leukocyte.        \\ \midrule
Binomial Options\cite{Podlozhnyuk2008:Binomial}   & Iteratively calculates the price for a portfolio of American stock options at multiple time points before expiration.  {\bfseries QOI:} The computed prices. \\ \midrule
MiniFE~\cite{minife}             & Proxy application for unstructured implicit finite element codes. {\bfseries QOI:} The final residual of the solver.                                           \\ \midrule
Black-scholes~\cite{Parsec}       & Analytically calculates the price for a portfolio of European stock options. {\bfseries QOI:} The computed prices.                                           \\ \midrule
LavaMD~\cite{Rodinia}             & Calculates particle potential and relocation due to forces between particles in a 3D space.  {\bfseries QOI:} The final force and location of each particle.  \\ \bottomrule
K-Means~\cite{Rodinia}            & Iterative clustering application that assigns observations to their closest cluster. {\bfseries QOI:} The cluster id each observation is assigned to.        \\ \midrule
\end{tabular}
\label{tab:benchmarks}
\end{table}

\begin{table}[]
\caption{Parameters used in our evaluation. Those listed under "Memo" are used for both memoization techniques. Only the AMD platform uses 64 tables per warp. \textit{small} and \textit{large} perforation use \textit{Items per Thread}.}
\label{tab:params}
\resizebox{\columnwidth}{!}{%
\begin{tabular}{@{}ll|ll@{}}
\toprule
\multicolumn{2}{l|}{\textbf{TAF}}                        & \multicolumn{2}{l}{\textbf{iACT}}                     \\ \midrule
\textit{hSize}                   & 1,2,3,4,5             & \textit{tPerWarp}         & 1,2,16,32,64              \\
\textit{pSize}                   & 2,4,8,\ldots,512      & \textit{tSize}            & 1,2,4,8                   \\
\textit{thresh}                  & 0.3,0.6,\ldots,1.5,3,5,20 & \textit{thresh}           & 0.1,0.3,\ldots,0.9,3,5,20 \\ \midrule
\multicolumn{2}{l|}{\textbf{perfo}}                      & \multicolumn{2}{l}{\textbf{Memo}}                     \\ \midrule
\textit{(large, small) skip}     & 2,4,8,16,32,64        & \textit{Hierarchy}        & thread, warp              \\
\textit{(ini, fini) skipPercent} & 10,20,\ldots,90       & \textit{Items per Thread} & 8,16,32,\ldots,512       
\end{tabular}%
}
\end{table}

\begin{figure*}
     \centering
        \includegraphics[width=0.99\textwidth, height=4.2cm]{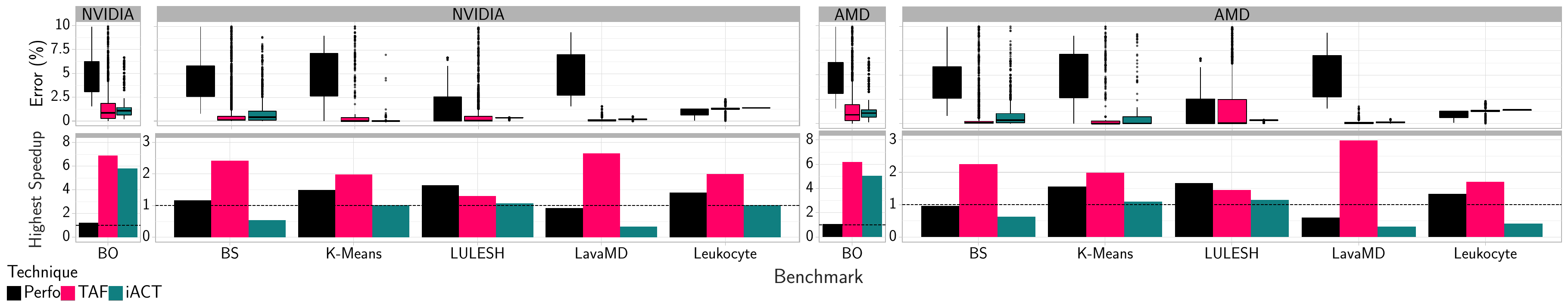}
        \caption{Highest speedup where error is less than 10\% (bottom) for perforation, TAF, and iACT for each benchmark on NVIDIA (left) and AMD (right). Also shown is the error distribution for all points where error is less than $10\%$ (top). We abbreviate Binomial Options with BO, and Blackscholes with BS. MiniFE is excluded because error is always greater than $10\%$. State-of-the-art AC techniques adapted to GPUs can significantly improve the performance of GPU-accelerated HPC applications.}
        \label{fig:summary}
\end{figure*}

\subsection{Benchmark Results}
Figure \ref{fig:summary} illustrates the highest speedup observed in our exploration for a maximum error of 10\% for both systems.
The figure depicts a clear trend: TAF approximate memoization is typically the best-performing approximation technique that meets
the imposed accuracy constraint, whereas input memoization performs worst. %

\begin{figure*}
    \centering
        \begin{subfigure}{0.30\linewidth}
        \centering
        \includegraphics[scale=0.35]{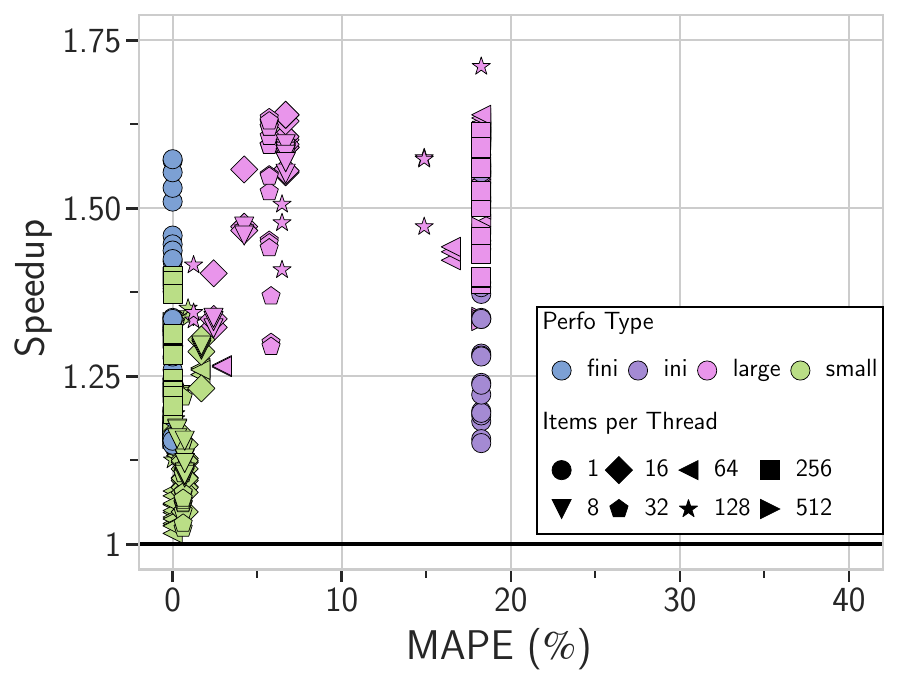}
        \caption{LULESH perfo results on NVIDIA.}
        \label{fig:lulesh_perfo_nvidia}
    \end{subfigure}
    \begin{subfigure}{0.30\linewidth}
        \centering
        \includegraphics[scale=0.35]{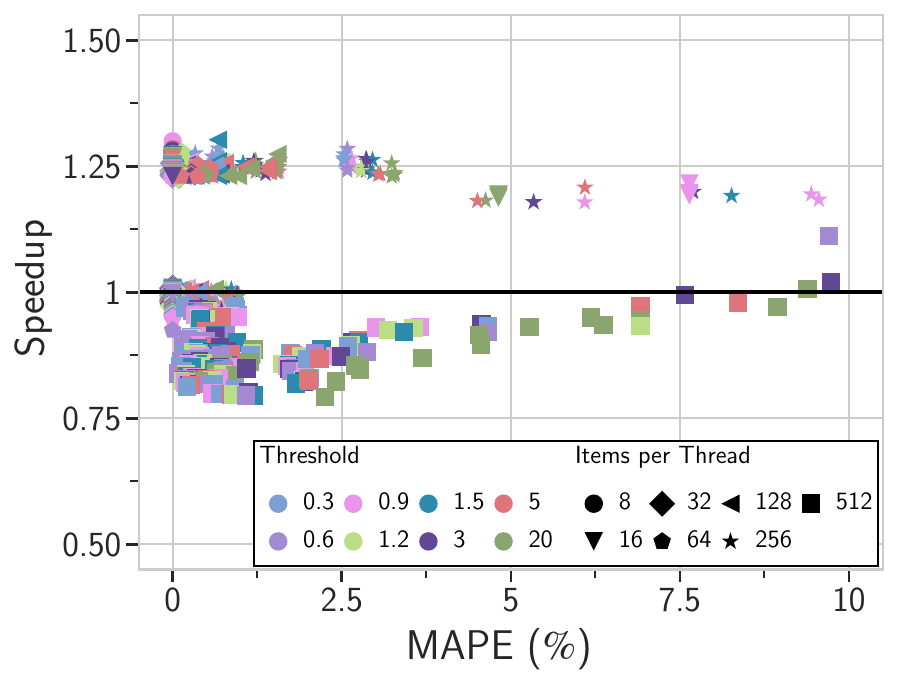}
        \caption{LULESH TAF results on NVIDIA.}
        \label{fig:lulesh_taf_nvidia}
    \end{subfigure}
    \hspace{1em}
    \begin{subfigure}{0.30\linewidth}
        \centering
        \includegraphics[scale=0.35]{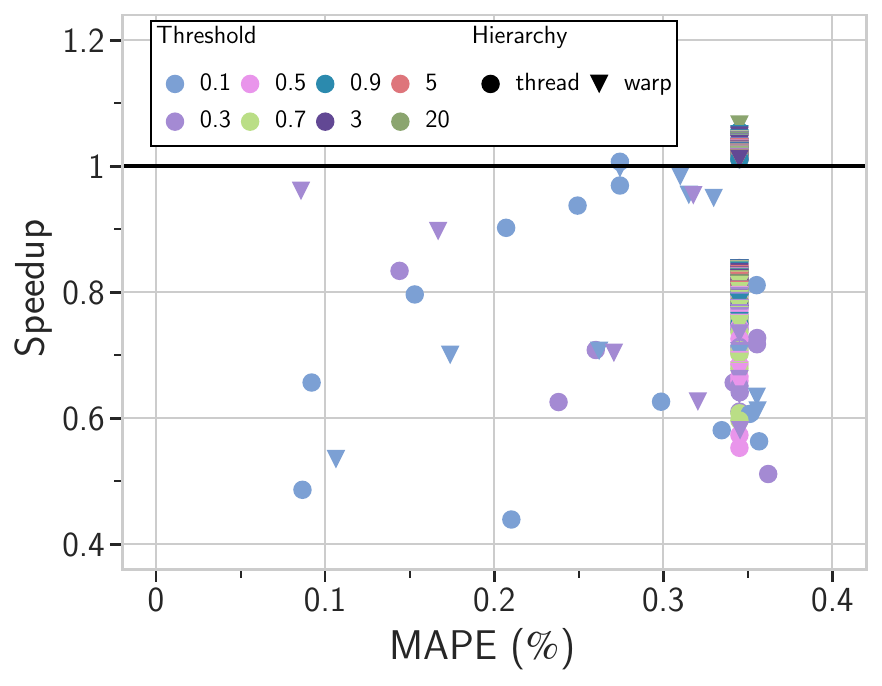}
        \caption{LULESH iACT results on NVIDIA.}
        \label{fig:lulesh_iact_nvidia}
    \end{subfigure}
    \hspace{1em}

    \begin{subfigure}{0.3\linewidth}
        \centering
        \includegraphics[scale=0.35]{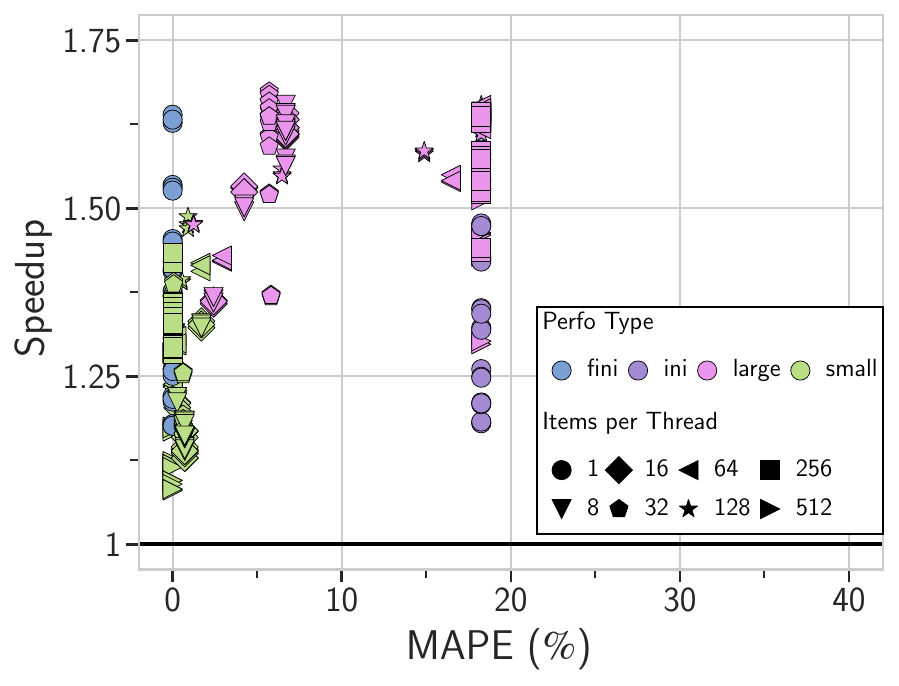}
        \caption{LULESH perfo results on AMD.}
        \label{fig:lulesh_perfo_amd}
    \end{subfigure}
    \hspace{1em}
    \begin{subfigure}{0.3\linewidth}
        \centering
        \includegraphics[scale=0.35]{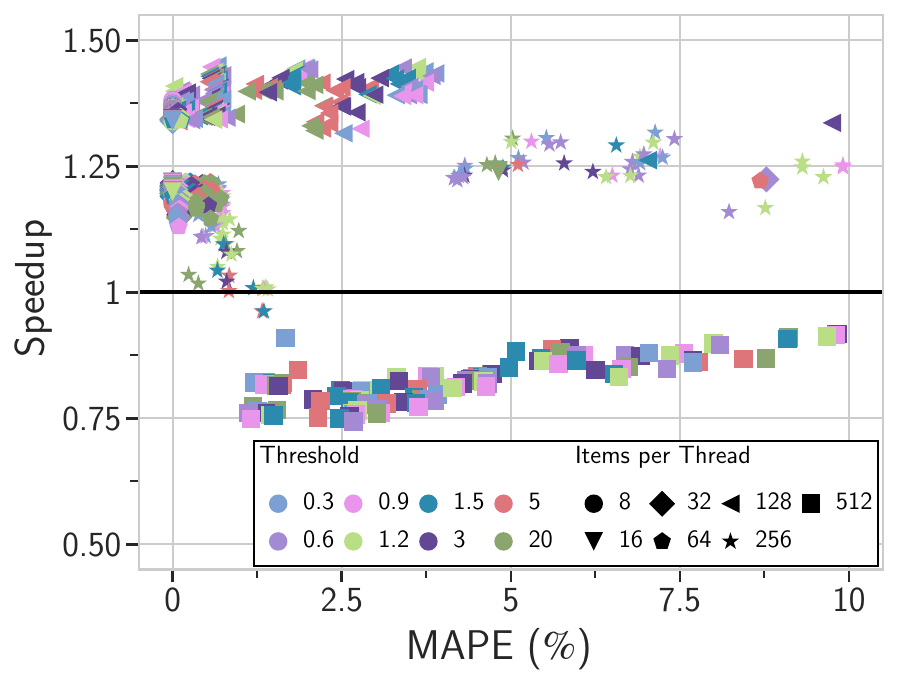}
        \caption{LULESH TAF results on AMD.}
        \label{fig:lulesh_taf_amd}
    \end{subfigure}
    \hspace{1em}
    \begin{subfigure}{0.3\linewidth}
        \centering
        \includegraphics[scale=0.35]{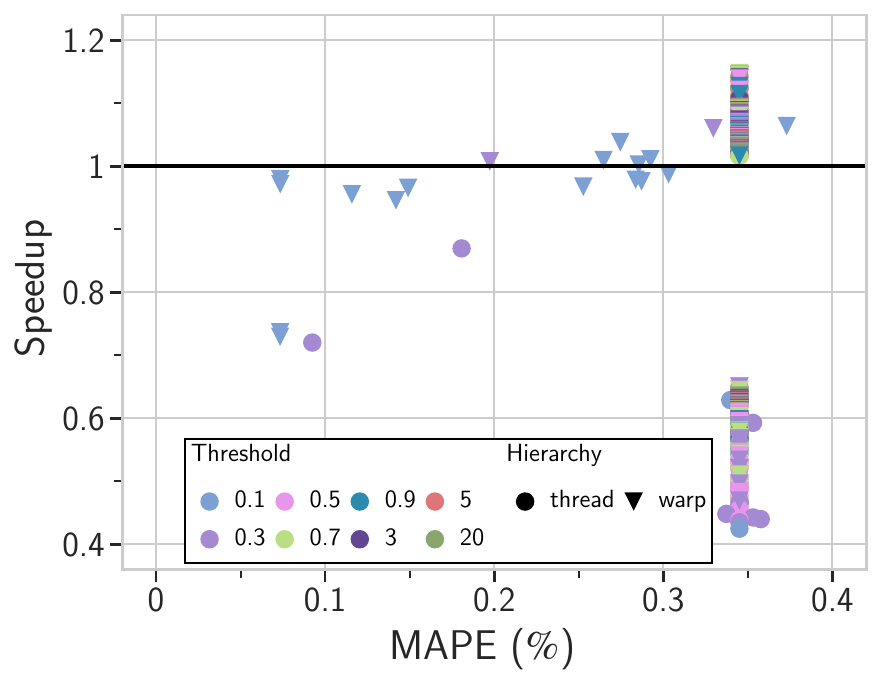}
        \caption{LULESH iACT results on AMD.}
        \label{fig:lulesh_iact_amd}
    \end{subfigure}

    \caption{LULESH results for TAF, iACT, and perforation on both platforms.}
    \label{fig:lulesh}
\end{figure*}

Figure ~\ref{fig:lulesh} depicts \textbf{LULESH} results for all approximation techniques on both systems.
We approximate the two most computationally expensive kernels: \verb|Calc|\verb|FBHour|\verb|glassForceForElems| and \verb|Calc|\verb|HourglassCont|\verb|rolForElems|.

Perforation accelerates LULESH by up to $1.64\times$ on NVIDIA and $1.67\times$ on AMD with less than $7\%$ MAPE.  \verb|fini| perforation induces less error than \verb|ini| perforation, indicating that the first iterations of the simulation contribute more to the output than the final ones.

Contrasting perforation, memoization techniques have less performance benefit but achieve much lower error. For TAF, we observe speedup
up to $1.30\times$ on NVIDIA, and up to $1.45\times$ on AMD with $0.67\%$ MAPE. iACT yields a lower error but cannot match TAF's performance. Specifically, iACT speedup is up to $1.07\times$ and up to $1.15\times$,
for NVIDIA and AMD, respectively with $0.3\%$ MAPE.

\begin{figure*}
    \begin{subfigure}{0.32\linewidth}
        \centering
        \includegraphics[scale=0.35]{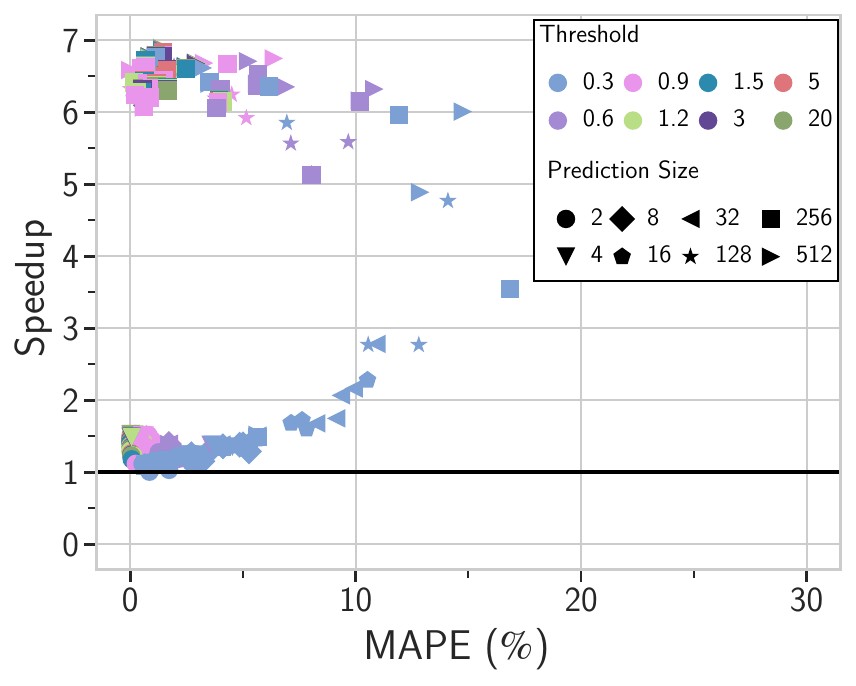}
        \caption{Binomial Options, TAF (NVIDIA).}
        \label{fig:bo_taf_nvidia}
    \end{subfigure}
    \begin{subfigure}{0.32\linewidth}
        \centering
        \includegraphics[scale=0.35]{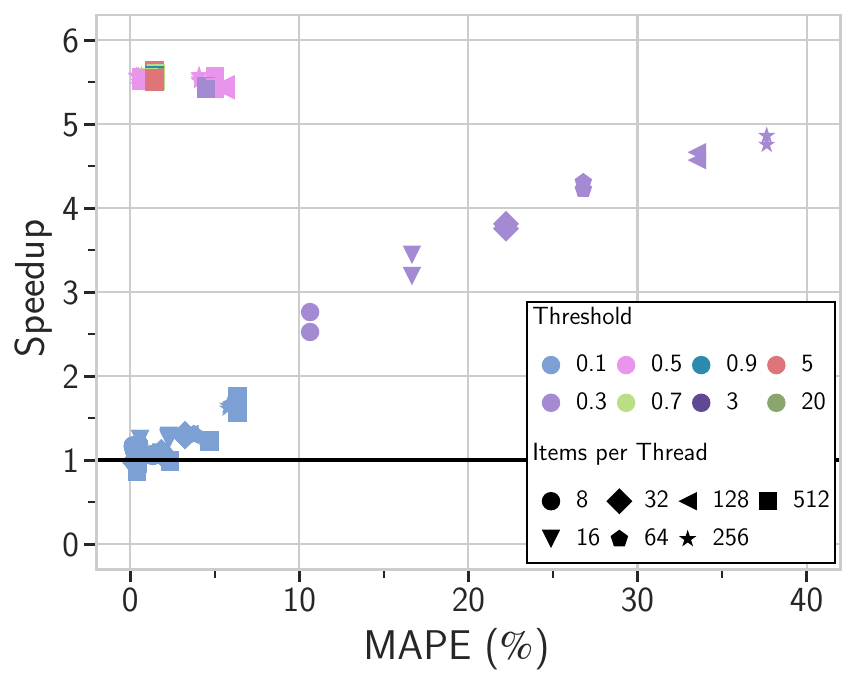}
        \caption{Binomial options, iACT (NVIDIA).}
        \label{fig:bo_iact_nvidia}
    \end{subfigure}
    \begin{subfigure}{0.32\linewidth}
     \centering
        \includegraphics[scale=0.35]{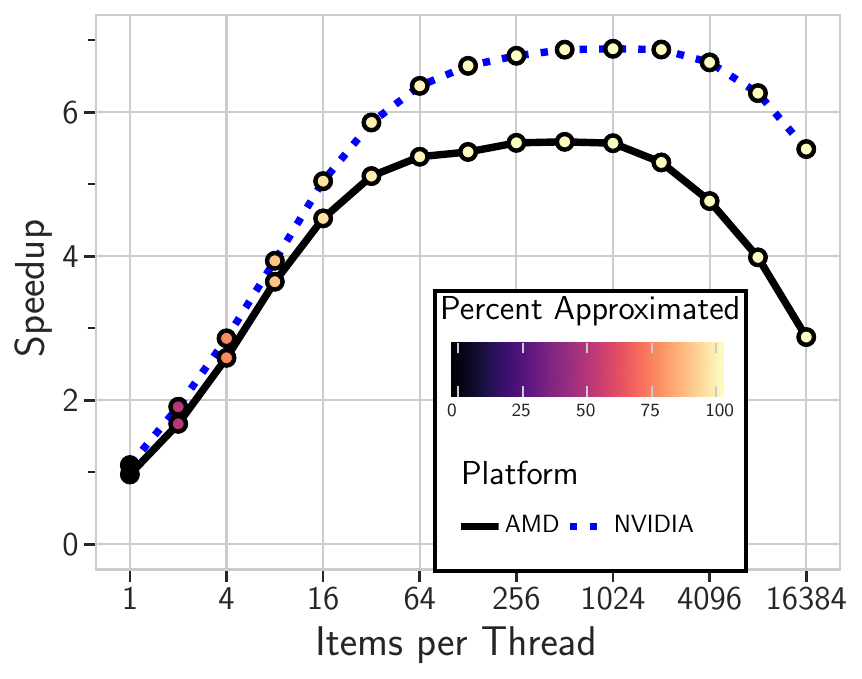}
        \caption{Parallelism versus approximation}
        \label{fig:parallelism_vs_approx}
    \end{subfigure}
    \caption{TAF and iACT results for Binomial Options (a), (d). In (c) the color scale indicates the percent of total price calculations that are approximated. Although the percent of approximated calculations approaches 100, speedup decreases as insufficient parallelism is available to hide latency.}
    \label{fig:res_bo}
\end{figure*}

In \textbf{Binomial Options}, an entire block collaboratively computes the price of a single option, and therefore we only use block-level decision-making. Both memoization techniques introduce minimal errors with
large performance benefits, indicating an ideal candidate for AC that demonstrates redundancy in the dataset which \hpacgpu{} can successfully exploit.
On the NVIDIA platform, TAF achieves up to $6.90\times$ 
speedup with $1.40\%$ MAPE (\ref{fig:bo_taf_nvidia}) while iACT achieves speedup up to $5.64\times$ speedup with $1.42\%$ MAPE (\ref{fig:bo_iact_nvidia}).

\begin{figure*}
    \centering
    \begin{subfigure}{0.30\linewidth}
        \centering
        \includegraphics[scale=0.33]{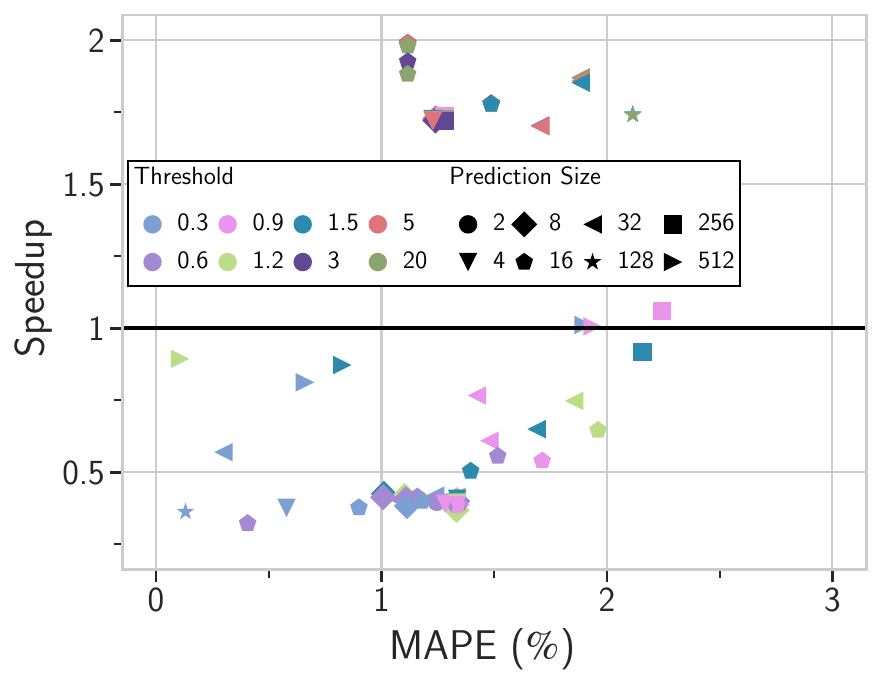}
        \caption{Leukocyte, TAF (NVIDIA).}
        \label{fig:leukocyte_taf_nvidia}
    \end{subfigure}
    \hspace{1em}
    \begin{subfigure}{0.30\linewidth}
        \centering
        \includegraphics[scale=0.33]{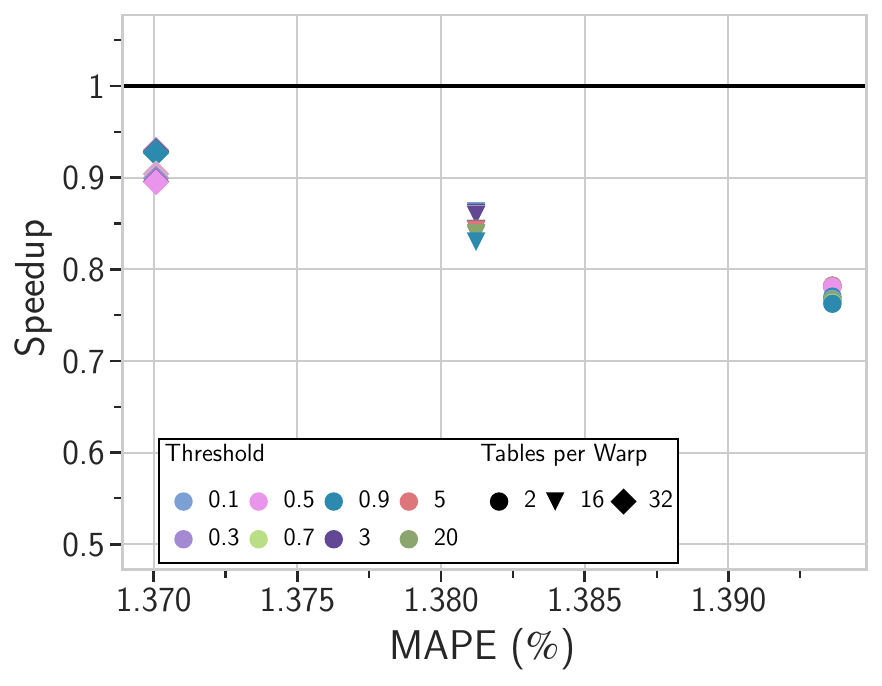}
        \caption{Leukocyte, iACT (NVIDIA).}
        \label{fig:leukocyte_iact_nvidia}
    \end{subfigure}
    \hspace{1em}
    \begin{subfigure}{0.30\linewidth}
        \centering
        \includegraphics[scale=0.33]{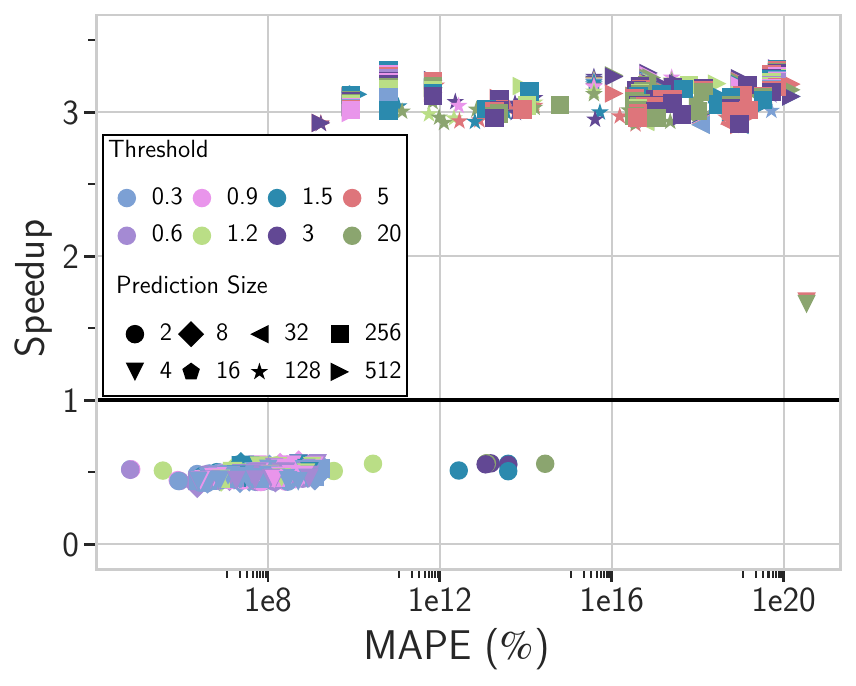}
        \caption{MiniFE, TAF (NVIDIA).}
        \label{fig:minife_taf_nvidia}
    \end{subfigure}
    \caption{TAF and iACT results for Leukocyte (a), (b),  and TAF for MiniFE on NVIDIA. iACT is not applicable to MiniFE.}
\end{figure*}

Approximate memoization depends on repeated executions of the same code region. For every execution, the approximate execution path may be activated, potentially increasing performance. OpenMP offload clauses allow developers to control the number of parallel threads that execute a region. Fewer threads results in more code region executions per thread, thereby increasing approximation potential, but more threads improve GPU latency hiding because there are more active warps. Thus, there is a conflict between parallel GPU execution and approximation benefits. For problem size $N$, approximation potential increases with the number of items computed by a thread. Maximizing approximation requires one thread to execute all $N$ iterations, whereas maximum parallelism requires $N$ threads, each executing one iteration.

Figure~\ref{fig:parallelism_vs_approx} demonstrates this tradeoff, showing the speedup from approximation for Binomial Options vs. the number of work items per thread.
On NVIDIA, speedup increases up to 2048 items per thread and then declines. The decline starts after 1024 items per thread on AMD. AMD performance decreases earlier because the AMD GPU has more SMs than the NVIDIA GPU, and thus needs more thread blocks executing to hide latency. At the same time, the percent of approximated calculations approaches 100. To achieve the best performance with approximation, it is important to balance approximation potential with the device's ability to hide latency.

In \textbf{Leukocyte}, we approximate the IMGVF matrix calculation and observe minimal quality reduction. TAF memoization (Figure~\ref{fig:leukocyte_taf_nvidia})
 results in speedups up to $1.99\times$ with an error of just $1.12\%$. 
In contrast to TAF, iACT reduces error but always slows down the application (Figure~\ref{fig:leukocyte_iact_nvidia}). The benefits of approximation are outweighed by the cost of the cache lookups and euclidean distance calculations in every invocation of the approximated function.  

In \textbf{MiniFE}, sparse matrix multiplication is approximated, resulting in locally introduced errors that propagate through subsequent iterations, causing high error rates (between $593\%$ and $3.43\times 10^{22}\%$, Figure~\ref{fig:minife_taf_nvidia}). iACT is not suitable since input sizes vary across threads due to the CSR matrix's non-zero values. \hpacgpu{} only supports computations with uniform input sizes for all threads.

\begin{figure*}
    \centering
    \begin{subfigure}{0.32\linewidth}
        \centering
        \includegraphics[scale=0.35]{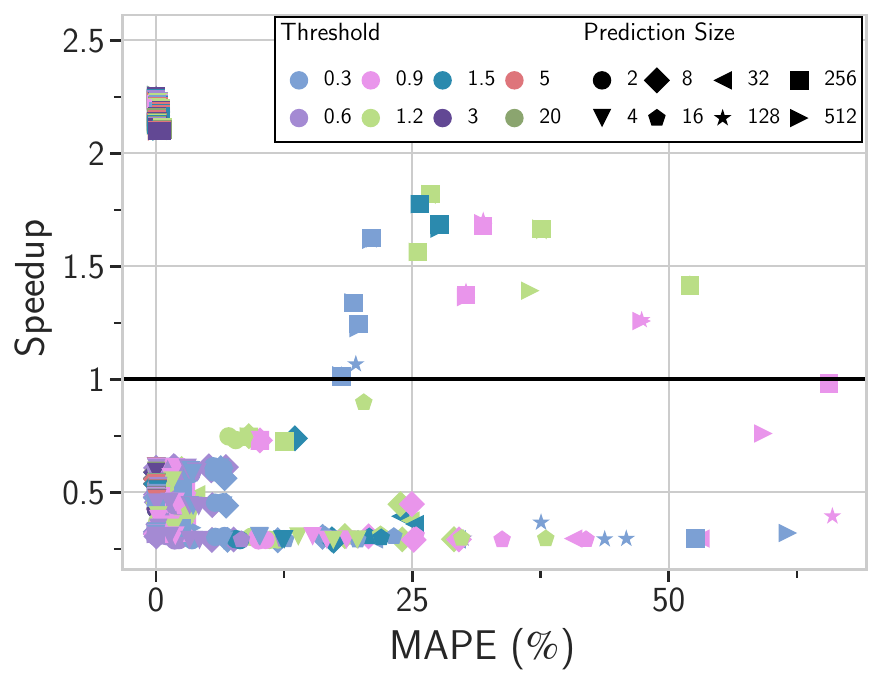}
        \caption{Blackscholes, TAF (AMD).}
        \label{fig:bs_taf_amd}
    \end{subfigure}
    \begin{subfigure}{0.32\linewidth}
        \centering
        \includegraphics[scale=0.35]{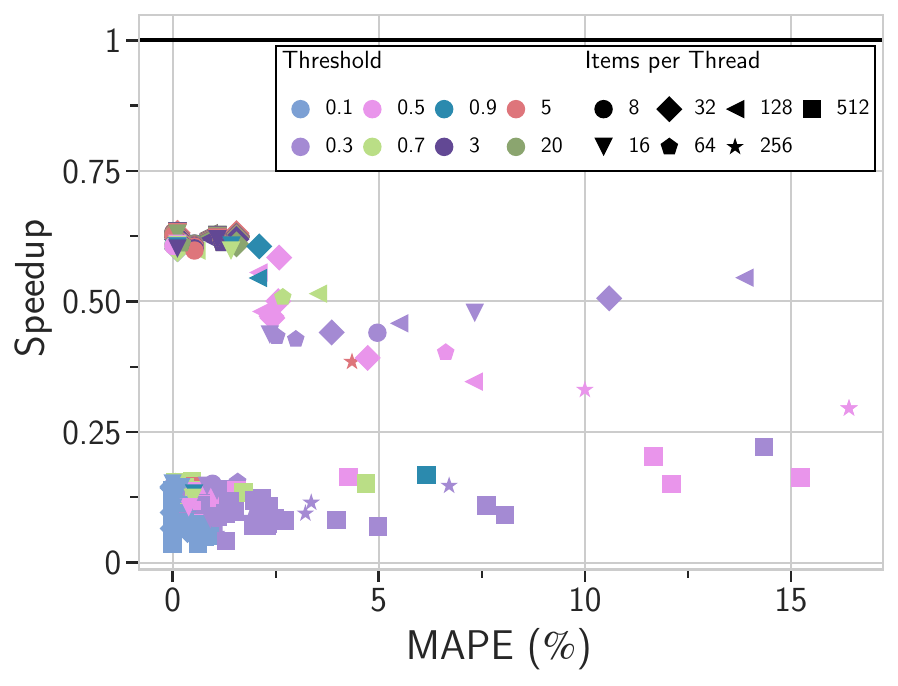}
        \caption{Blackscholes, iACT (AMD).}
        \label{fig:bs_iact_amd}
    \end{subfigure}
    \begin{subfigure}{0.32\linewidth}
         \centering
         \includegraphics[scale=0.35]{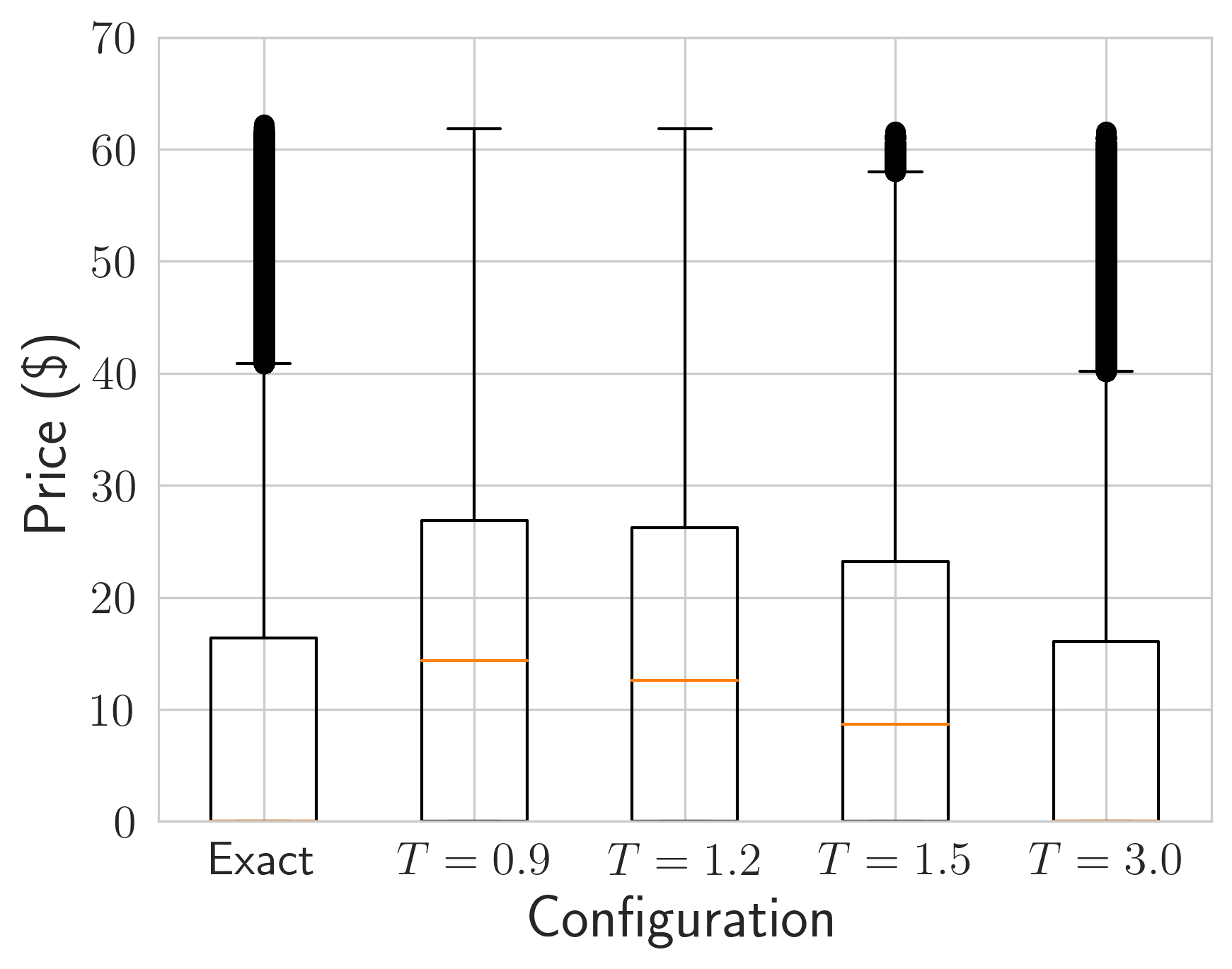} 
        \caption{Distributions of outputs.}
        \label{fig:taf_bs_threshold}
    \end{subfigure}
    \caption{Results of blackscholes on the AMD system. In (c) the distribution of the Exact output prices vs. prices calculated using TAF with history size $5$, prediction size $512$, and RSD threshold $T$. When the RSD threshold is too low, unrepresentative values are output and the error is high. TAF RSD interacts with the application to produce unintuitive results.}
    \label{fig:res_blck} 
\end{figure*}
In \textbf{Blackscholes}, $99\%$ of the time is spent in memory allocations and data transfers between the host and device. However, we 
approximate the device kernel (the entire price calculation of an option).
Therefore, we present kernel performance only.
TAF is extremely effective with speedups of up to $2.26\times$ on the
AMD platform with $0.015\%$ MAPE (Figure~\ref{fig:bs_taf_amd}).
Performance is best when prediction size and threshold are high,
suggesting the aggregate output table across all threads represents the data well.

Blackscholes demonstrates the weaknesses in TAF's activation criterion (RSD). One would expect application error and performance to increase with RSD, but Figure~\ref{fig:bs_taf_amd} shows otherwise. Figure~\ref{fig:taf_bs_threshold} compares prices from Blackscholes using TAF to the original benchmark at different threshold values. For thresholds below $3.0$, approximation is activated, and the error is high. At $T=3.0$, the approximated values closely match the original, yielding low error.

\begin{figure*}[t]
    \hspace{1em}
    \begin{subfigure}{0.32\linewidth}
        \centering
        \includegraphics[scale=0.42]{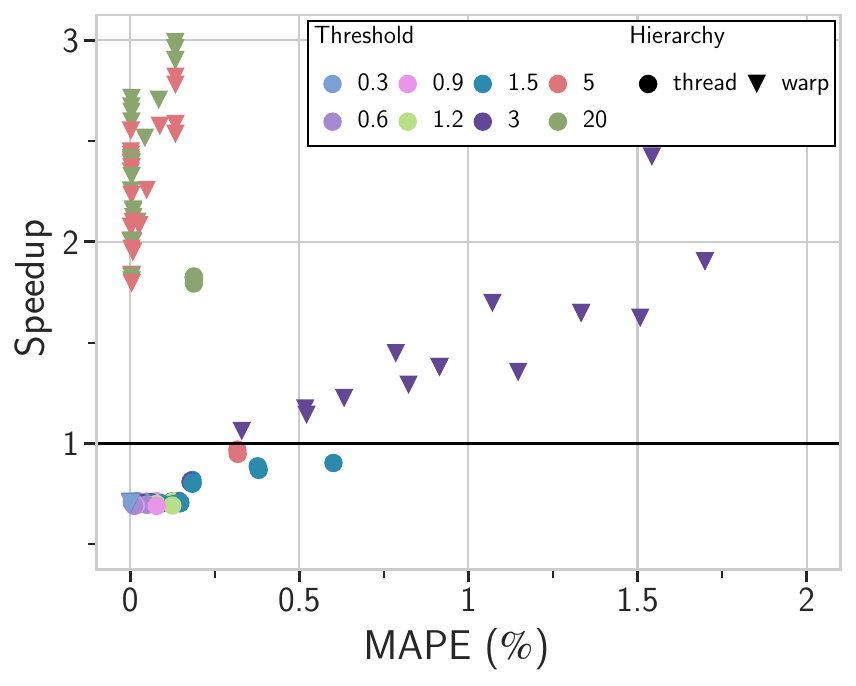}
        \caption{LavaMD, TAF (AMD).}
        \label{fig:lavamd_taf_amd}
    \end{subfigure}
    \begin{subfigure}{0.32\linewidth}
        \centering
        \includegraphics[scale=0.42]{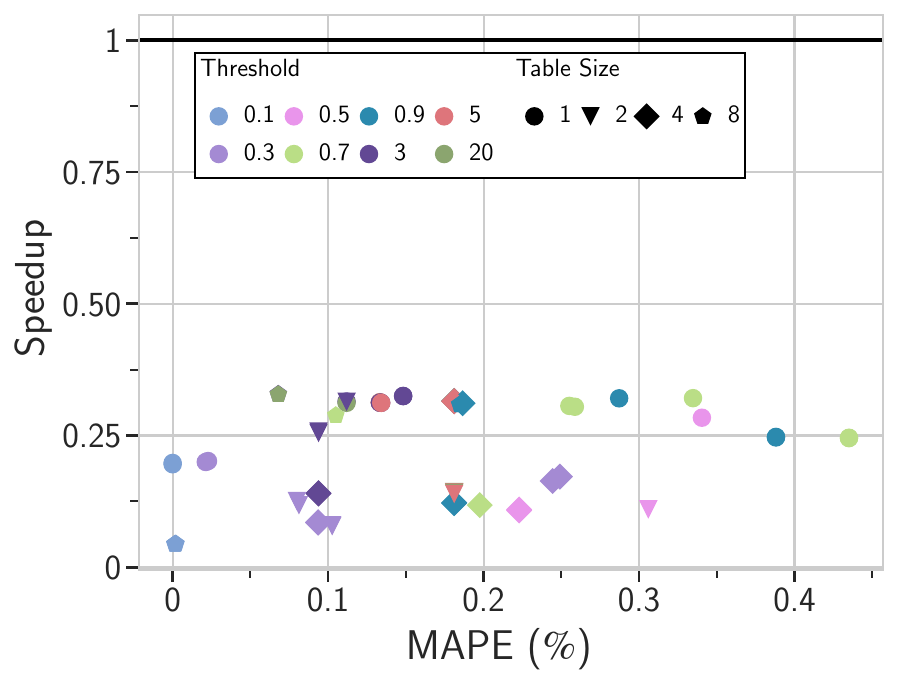}
        \caption{LavaMD, iACT (AMD).}
        \label{fig:lavamd_iact_amd}
    \end{subfigure}
    \begin{subfigure}{0.32\linewidth}
    \centering
    \includegraphics[scale=0.42]{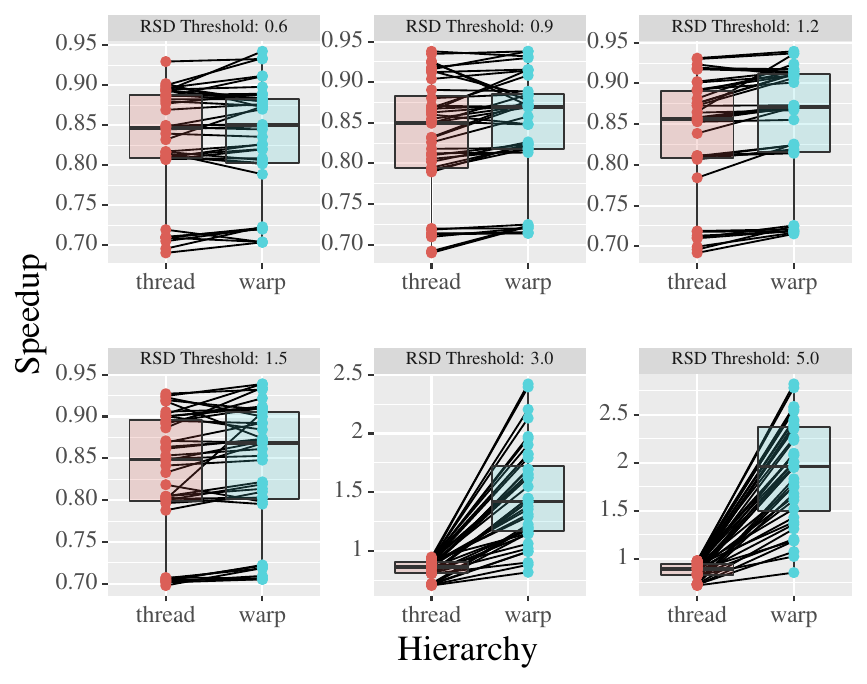}
        \caption{}
        \label{fig:lavamd_hierarchy}
    \end{subfigure}
    \label{fig:res_lavaMD}
    \caption{LavaMD: results when approximated using TAF (a) and iACT(b). In (c) the paired line boxplot depicts the speedup on AMD. Lines connect points whose configurations differ only in decision hierarchy, and the boxplots show the speedup distribution for a given threshold and hierarchy level. Warp-level can increase speedup by eliminating control divergence.}
\end{figure*}

In \textbf{LavaMD}, the force calculation for neighboring boxes is approximated. As shown in Figure~\ref{fig:lavamd_taf_amd}, TAF offers significant performance gains ($2.98\times$ speedup) with minimal errors ($0.133\%$). Higher RSD thresholds and prediction sizes yield better results, similar to Blackscholes. Figure~\ref{fig:lavamd_iact_amd} indicates that iACT has lower error than TAF but slows down the application. This is due to the higher cost of accessing the shared cache table and computing euclidean distances, compared to the original force computation.

\begin{figure*}[t]
    \begin{subfigure}{0.32\linewidth}
        \centering
        \includegraphics[scale=0.37]{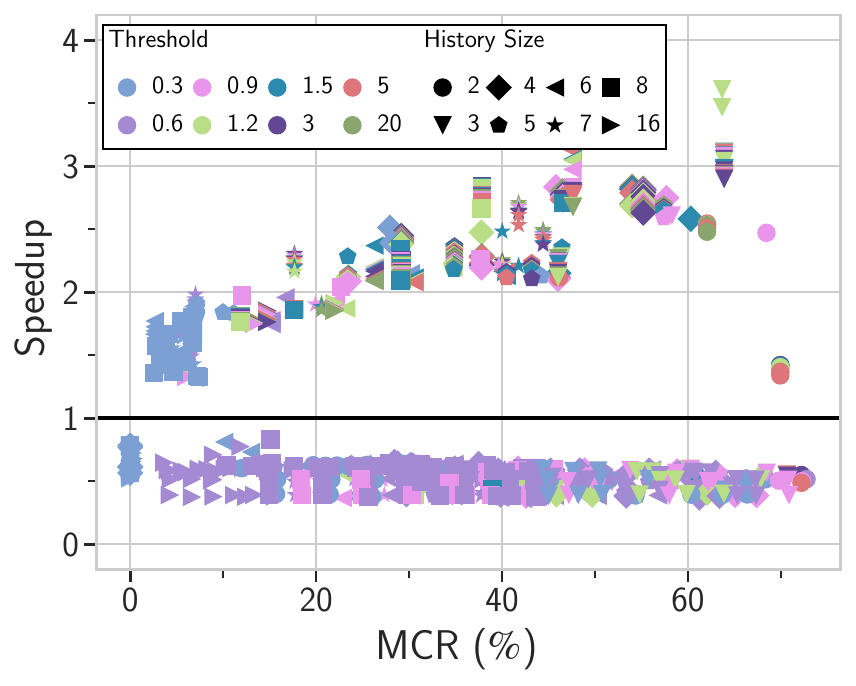}
        \caption{TAF (AMD).}
        \label{fig:kmeans_taf_amd}
    \end{subfigure}
    \hspace{1em}
    \begin{subfigure}{0.32\linewidth}
        \centering
        \includegraphics[scale=0.37]{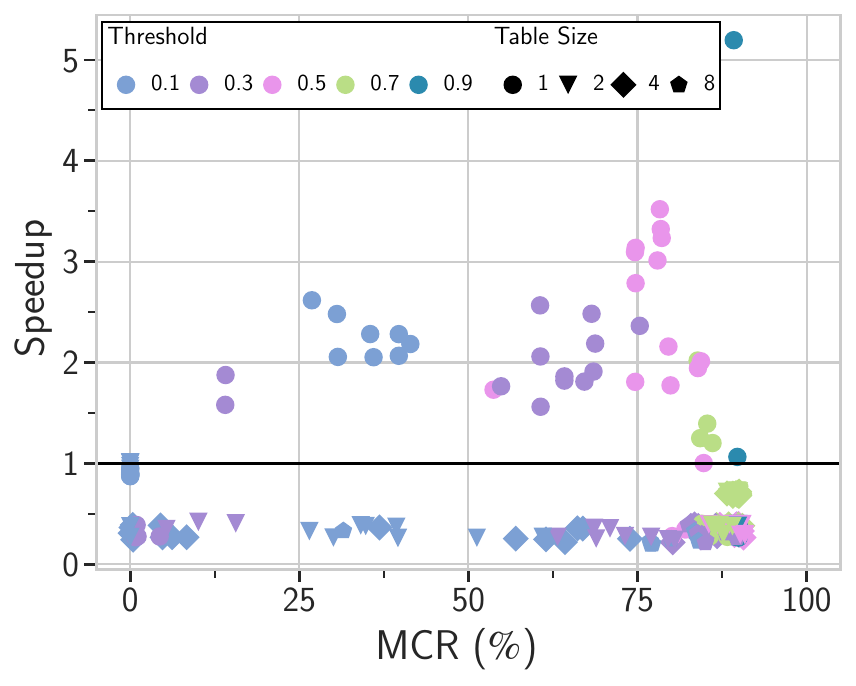}
        \caption{iACT (AMD).}
        \label{fig:kmeans_iact_amd}
    \end{subfigure}
    \begin{subfigure}{0.32\linewidth}
    \centering
    \includegraphics[scale=0.37]{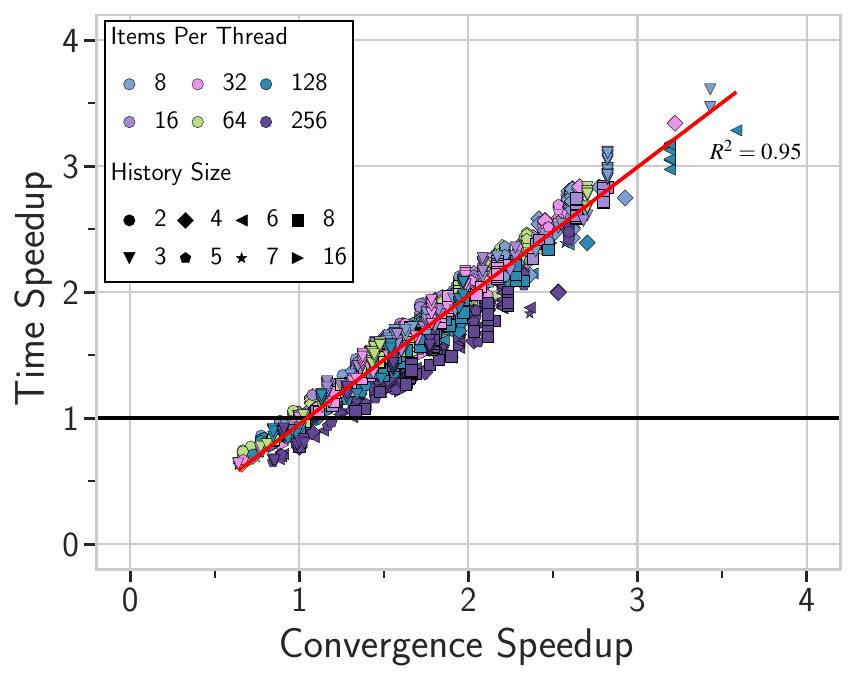}
    \caption{Convergence speedup vs time speedup}
    \label{fig:kmeans_convergence}
    \end{subfigure}
    \caption{K-Means (a) TAF and (b) iACT results for K-Means. (c) In K-means, speedup comes primarily from early convergence. Time Speedup (y-axis) vs. Convergence Speedup (x-axis), where if the non-approximated kernel converges in $n$ iterations and the approximated kernel converges in $a$ iterations, then convergence speedup is: $n/a$}
    \label{fig:res_kmeans}
\end{figure*}
Warp-level decision-making can increase performance. In Figure~\ref{fig:lavamd_hierarchy}, we contrast the speedup obtained in LavaMD by thread-level and warp-level decision-making
for different RSD thresholds. Warp-level decision-making, for a given threshold, increases the median speedup by up to $2.27\times$. 
 In warp-level decision-making, threads decide as a group whether
to approximate, eliminating approximation-induced thread divergence. However,
divergence elimination is not the only cause of performance benefits. Warp-level decision-making can force threads
with RSD values larger than the threshold to approximate because they are
within the minority of the group. In such cases, \hpacgpu{} increases approximation.

\textbf{K-Means}.
In K-Means, we approximate the kernel computing the euclidean distance of observations with the current clusters. Although 
this kernel accounts for $3.5\%$ of the total benchmark execution time, we observe substantial performance benefits ($3.5\times$ speedup). The introduced inaccuracies cause observations to stay in the same cluster, affecting the convergence criterion. K-Means converges when no observations change cluster. Observations are herded to the same cluster by memoization techniques that use previous computations. Figure\ref{fig:kmeans_convergence} indicates a strong linear correlation ($R^2=0.95$) between the observed speedup and
the speedup obtained by converging early. Figures~\ref{fig:kmeans_taf_amd}, \ref{fig:kmeans_iact_amd} summarize the performance accuracy 
trade-offs of K-Means using TAF and iACT.

\subsection{\hpacgpu{} limitations}
Approximation speedup and error are influenced by code region sensitivity and input data, which can vary significantly across applications and the situation in which they are applied. In this work, we study methods and algorithms for porting state-of-the-art AC techniques to GPUs without investigating how different datasets affect error. Given representative input data, the \hpacgpu{} execution harness allows exploring whether AC yields tolerable error for a given application and input data set. 

The user is faced with a large parameter space over which the \hpacgpu{} execution harness performs an exhaustive search, costing compute time\footnote{All experiments for a given benchmark took up to $988$ GPU hours (median $1.35$ hours).} to perform the experiments, and costing end-user time to process the results and decide which code regions to approximate. Given this cost, we believe there is considerable value in work that automates the end-to-end workflow. Such automation could integrate with sensitivity analysis tools~\cite{Vassiliadis2016:Towards,Parasyris2022:Approximate,Menon2018SC} to find code regions amenable to approximation and smart search/optimization techniques (genetic algorithms, Bayesian Optimization) to reduce parameter exploration costs.

Despite these limitations, \hpacgpu{}'s prescriptive programming model enables our extensive analysis of general-purpose approximate computing for GPUs in HPC to derive deep insights that can guide current use and future systems.
\subsection{Insights}
Our analysis conducted with $57,288$ configurations is, to our knowledge, the most comprehensive study of GPU-accelerated AC ever performed. This analysis and our experience approximating HPC-based applications on the GPU yield several insights into the interplay between approximate computing and GPU-based parallelism:%
\begin{enumerate}[leftmargin=*]
    \item State-of-the-art approximation techniques adapted to GPU architectures can significantly improve the performance of GPU-accelerated HPC applications (Figure~\ref{fig:summary}). Each application has unique, unpredictable performance-accuracy trade-offs.
    \item Approximation techniques accelerate applications while reducing available parallelism (Figure~\ref{fig:parallelism_vs_approx}). This trade-off is critical as AC affects the GPU's ability to hide latency. Speedup for TAF and iACT decreases as the number of SMs in the GPU increases.
    \item TAF RSD behaves differently in each application in ways that can produce unintuitive results (Figures~\ref{fig:bo_taf_nvidia},\ref{fig:bs_taf_amd},~\ref{fig:taf_bs_threshold}). Predictable interactions between approximation and speedup/quality loss are important for AC use in production codes.
    \item TAF has higher speedup than iACT (Figures~\ref{fig:res_bo}-\ref{fig:res_kmeans}). iACT must always pay the cost of deciding whether to approximate, while TAF can amortize the decision-making cost by approximating multiple times.
    \item Load imbalance caused by control divergence within a warp can degrade performance in GPU-based AC (Figure~\ref{fig:lavamd_hierarchy}). Hierarchical decision-making improves performance by eliminating control divergence. 
    \item iACT is slower than TAF but introduces less error, suggesting that matching based on euclidean distance is more suitable for applications than RSD values. Similar to custom hardware to accelerate machine learning, hardware support for approximate memoization tables can help realize the accuracy benefits of iACT with improved performance.
\end{enumerate}

\section{Related Work}
\label{sec:related}
Approximate Computing has gained significant attention in recent years as a promising approach for improving performance and energy efficiency. There are numerous strategies from lower-level hardware techniques, such as approximate floating-point multipliers~\cite{froehlich2018towards, Rehman2016,kulkarni2011trading}, inexact adders~\cite{kahng2012accuracy, gupta2011impact}, voltage scaling~\cite{chippa2014scalable, esmaeilzadeh2012architecture,hedge1999energy,sampson2011enerj}, to higher-level software techniques including loop perforation \cite{Hoffmann2009, goiri2015approxhadoop}, function memoization\cite{michie1968memo, keramidas2015clumsy, samadi2014paraprox}, mixed-precision~\cite{Kotipalli2019AMPTGA,Menon2018SC,Laguna2019GPUMixer,gonzalez2013precimonious,Gu2020:FPTuner,Kang2020:PreScaler} etc.
Our work is complementary to those studies as we introduce a new framework that enables the use and evaluation of approximate techniques on GPUs.

Several frameworks facilitate approximate computing on the CPU. For example, GREEN~\cite{baek2010green} is an API that enables loop and function-level approximations and generates functions for application-specific error models. 
ACCEPT~\cite{sampson2015accept} is a compiler framework developed on top of the LLVM compiler infrastructure. It automatically selects the best approximation strategy for a given quality threshold provided via programmer annotations. In~\cite{Vassiliadis2015:Programming}, authors create a task-based programming model allowing users to specify task significance and provide approximate versions of tasks. 
ApproxHPVM~\cite{sharif2019approxhpvm} uses an intermediate representation to enable accuracy-aware compiler optimizations for ML applications. The authors in ~\cite{sharif2021approxtuner} extend ApproxHPVM with an auto-tuning framework for approximate aware optimizations of tensor-based applications.
HPAC~\cite{parasyris2021hpac} is a framework that supports approximations in OpenMP applications and studies the interaction between approximate techniques and parallelism. While these works cater to CPU applications, we focus on GPU kernel approximations. %

There have been frameworks designed for GPU applications.
SAGE~\cite{Samadi2013} is a compiler and a runtime system that targets machine learning and image processing applications. It automatically generates approximate kernels for GPUs and selects suitable kernels to approximate at runtime to meet a user-provided quality threshold.
TruLook~\cite{garcia2021trulook} is a configurable approximation framework that accelerates GPU applications. Using an approximate multiplier, it supports computational reuse via a lookup table and approximate arithmetic operations. While TruLook provides hardware support for approximation, our framework is designed for software approximation techniques.
Authors of \cite{koutsovasilis2018achee} evaluated several applications on CPUs and GPUs, studying the cumulative effect of heterogeneity and approximations on energy footprint and output quality. Our work, on the other hand, is focused on improving performance.

Besides those frameworks that provide programmatic support for approximation, others help users decide where to approximate. The ASAC~\cite{Roy2014:ASAC} framework automatically identifies approximable data in a program via sensitivity analysis. Authors of~\cite{Vassiliadis2016:Towards} employ profiling and interval analysis to create a programming model that finds code amenable to approximation. Puppeteer~\cite{Parasyris2022:Approximate} measures code region sensitivity using uncertainty quantification, and ranks the regions according to their sensitivity.

\section{Conclusion}\label{sec:conclusion}
This paper introduces \hpacgpu, a portable pragma-based programming model that extends OpenMP offload applications to support Approximate Computing techniques in GPU-based HPC systems with minimal code changes. Our comprehensive analysis demonstrates that state-of-the-art approximation techniques, when adapted to fit the distinct architectural features of GPUs, can significantly improve the performance of GPU-accelerated HPC applications with minimal quality loss. We demonstrate that \hpacgpu achieves up to $6.9\times$ speedup on HPC-centric workloads while introducing less than $10\%$ error.

The insights provided in this paper contribute to a deeper understanding of the interplay between approximate computing and GPU-based parallelism, guiding the future development of AC algorithms and systems for these architectures. For instance, hierarchical decision-making can help eliminate control divergence. As the end of Dennard scaling and the slowdown of Moore's law challenge the advancement of computing performance, approximate computing in HPC systems offers a promising avenue to continue pushing the boundaries of performance.

\begin{acks}
The authors would like to thank the anonymous reviewers for
their valuable comments and helpful suggestions. 

The views and opinions of the authors do not necessarily reflect those of the U.S. government or Lawrence Livermore National Security, LLC neither of whom nor any of their employees make any endorsements, express or implied warranties or representations or assume any legal liability or responsibility for the accuracy, completeness, or usefulness of the information contained herein. This work was prepared by LLNL under Contract DE-AC52-07NA27344 (LLNL-CONF-847178) and was supported by both the LLNL LDRD Program under Projects No. 20-ERD-043 and 21-ERD-018.

\end{acks}

\bibliographystyle{ACM-Reference-Format}
\bibliography{ref.bib}

\appendix

\end{document}